\documentclass[aps,preprint,showkeys]{revtex4-1}
\usepackage{amsmath}
\usepackage{amssymb}
\usepackage{graphicx}
\usepackage{enumerate}

\numberwithin{equation}{section}% reset equation counter for sections

\usepackage{epstopdf}

\begin{document}

\title[Existence of optical vortices in saturable nonlinear media]{On the existence of optical vortex solitons propagating in saturable nonlinear media}

\author{Luciano Medina}
\email{lmedina@nyu.edu}
\affiliation{Department of Mathematics\\ 
Tandon School of Engineering\\ 
New York University\\
Brooklyn, NY 11201
USA}

\date{\today}

\newtheorem{theorem}{Theorem}
\newtheorem{lemma}[theorem]{Lemma}
\newtheorem{prop}[theorem]{Proposition}

\numberwithin{theorem}{section}
\numberwithin{equation}{section}

\begin{abstract}
In this paper, an existence theory is established for ring-profiled optical vortex solitons. We consider such solitons in the context of an electromagnetic light wave propagating in a self-focusing nonlinear media and governed by a nonlinear Schr\"odinger type equation. A variational principle and constrained minimization approach is used to prove the existence of positive solutions for an undetermined wave propagation constant. We provide a series of explicit estimates related to the wave propagation constant, a prescribed energy flux, and vortex winding number. Further, on a Nehari manifold, the existence of positive solutions for a wide range of parameter values is proved. We also provide numerical analysis to illustrate the behavior of the soliton's amplitude and wave propagation constant with respect to a prescribed energy flux and vortex winding number.
\end{abstract}
\maketitle

\section{Introduction}

An exciting area of research in modern optics is the study of optical vortices. In a light wave, optical vortices are formed by wave dislocations or defects \cite{NB}. Its applications are found across numerous branches of nonlinear science, such as quantum information processing, wireless communications, and some not directly related to wave propagation, for instance condensed matter physics, particle interactions, and cosmology  \cite{ABSW,BKKH,BSV,DKT,KMT,RSS,SO, SL,YW}. An interesting class of optical vortices are the ring-profiled optical vortices. Such vortices can be considered as a ring of light with a black spot at its center. In terms of a light beam, such black spots represent a zero light intensity.

Of particular interest, is the theoretical description of a complex-valued light wave propagating in a nonlinear media and governed by a nonlinear Schr\"odinger equation \cite{A,DY,KVT1,KVT2,MSZ,NAO,RLS,SF}. Rigorous mathematical treatments of such nonlinear problems present mathematical challenges and have been considered by mathematical analysts \cite{AC, AMN,ASY,BW,BZ,CL,LiuRen, Sch,YZ,YZ2}. Our interest is motivated by the work of  Skryabin and Firth \cite{SF} and the mathematical analysis of Liu and Ren \cite{LiuRen}, and Zhang and Yang \cite{YZ2}.

Consider the propagation of an electromagnetic wave in the longitudinal $z$-direction over the transverse plane of coordinates $(x,y)$ perpendicular to the $z$-axis. In dimensionless form, the evolution of the slowly varying electric field envelope, $E$, is modelled by the nonlinear Schr\"odinger equation \cite{SF},
\begin{equation}
i\frac{\partial E}{\partial z}+\dfrac{1}{2}\nabla^2_\perp E+F(I)E=0,\label{NLSeq}
\end{equation}
where $\nabla^2_\perp$ is the Laplace operator over the transverse plane of coordinates. The function $F$ depends on the total field intensity, $I$, i.e., $I=|E|^2$, and encapsulates the nonlinear properties of an optical medium. Examples for $F(I)$ include
\begin{enumerate}
\item $F(I)= I$ (pure Kerr nonlinearity),
\item $F(I)= I-\alpha I^2$ (cubic-quintic model),
\item $F(I)= I(1+\alpha I)^{-1}$ (saturable nonlinearity),
\end{enumerate}
where $\alpha\in\mathbb{R}$ is a parameter describing the nonlinearity saturation \cite{DKT}. The saturation constant may be defined as $\alpha=l_{nl}/l_{d}$, with $l_{nl}$ and $l_d$ denoting the nonlinear and diffraction lengths, respectively. Note that the saturable nonlinearity and cubic-quintic models reduce to the pure-Kerr nonlinearity when the saturation constant is zero. 

We focus on spatial optical solitons \cite{DKT, KS, S, SS}. Spatially localized solutions of \eqref{NLSeq}, which do not change their intensity profile during propagation, can be described under the spatial soliton ansatz 
\begin{equation}
E(x,y,z)=u(x,y)\exp(i\kappa z+i\phi(x,y)),\label{SpatialAnsatz}
\end{equation}
where $u$ and $\phi$ are real valued functions representing the soliton amplitude and phase, respectively, and $\kappa\in\mathbb{R}$ is the wave propagation constant. In view of \eqref{SpatialAnsatz}, the nonlinear Schr\"odinger equation \eqref{NLSeq} transforms into the coupled system
\begin{align}
\left\{\begin{array}{c}
\nabla^2_\perp u-|\nabla\phi|^2u-2[\kappa-F(u^2)]u=0,\\
\nabla^2_\perp\phi+2\nabla\phi\cdot\nabla\ln(u)=0.
\end{array}\right.\label{coupledSys}
\end{align}
Under an appropriate balance between the nonlinear and diffraction lengths, the electromagnetic radiation may become self-trapped and form a self-induced wave-guide. Solutions to \eqref{coupledSys} are referred to as self-trapped nondiffracting solutions in a self-focusing saturable nonlinearity.

The existence theory in this paper seeks self-trapped positive radially symmetric solutions of \eqref{coupledSys} with a phase singularity at its center. Such solutions describe ring-profiled vortex solitons and can be found under the $n$-vortex ansatz 
\begin{equation}
u=u(r),\quad \phi=n\theta,\quad r=\sqrt{x^2+y^2},\quad \theta=\arctan(y/x),\label{vortexAnsatz}
\end{equation}
where $r,\theta$ are polar coordinates over $\mathbb{R}^2$ and $n\in\mathbb{Z}$ is the vortex winding number. 

Due to the presence of the vortex core or, equivalently, the regularity of $u$ at $r=0$, we impose the condition $u(0)=0$. Moreover, such ring-like beams remain localized.  Thus allowing us to mathematically impose the ``boundary'' condition $u(R)=0$ for $R>0$ sufficiently large, where $R$ represents the distance from the vortex core. 

Using \eqref{vortexAnsatz}, and in a saturable nonlinear media, the system \eqref{coupledSys} reduces to the $n$-vortex equation
\begin{align}
\left\{\begin{array}{c}
(ru_r)_r-\dfrac{n^2}{r}u+\dfrac{2ru^3}{1+\alpha u^2}-2\kappa ru=0,\\
u(0)=0, \qquad u(R)=0.
\end{array}\right.\label{vortexEq}
\end{align}
An important parameter characterization of spatial solitons is its energy flux. Using the $n$-vortex ansatz, the soliton energy flux is defined as
\begin{equation}
Q(u)=\int_{0}^{R}|E|^2rdrd\theta=2\pi\int_{0}^{R}ru^2dr.\label{energyFlux}
\end{equation}

The rest of the paper is summarized as follows. In section 2, we give a necessary condition for the existence of nontrivial solutions. In section 3, we treat \eqref{vortexEq} as a nonlinear eigenvalue problem and use a constrained minimization approach, subject to a prescribed energy flux constraint, to prove the existence of positive solution pairs $(u,\kappa)$. In section 4, we prove the existence of positive solutions for a wide range of parameter values over a Nehari manifold. In section 5, we supplement our results by using a finite element formalism to compute the soliton's amplitude and wave propagation constant for a prescribed energy flux. A summary is provided in section 6. 

% % % % % % % % % % % % % % % % % % % % % % % % % % % % % % % % % %
% % % % % % % % % % % % % % % % % % % % % % % % % % % % % % % % % %
\section{Necessary Condition for nontrivial solutions}
The $n$-vortex equation \eqref{vortexEq} may be viewed as the Euler-Lagrange equations of a corresponding action functional. For a sufficiently large distance $R$, we prove that the action functional is indefinite, and, therefore, a direct minimization approach is not possible.

Consider the action functional $\mathcal{I}_{\kappa}:H\rightarrow\mathbb{R}$ defined as
\begin{equation}
\mathcal{I}_{\kappa}(u)=\frac{1}{2}\int_{0}^{R}\left\{ru^2_r+\frac{n^2}{r}u^2-2(\alpha^{-1}-\kappa)ru^2+2\alpha^{-2}r\ln(1+\alpha u^2)\right\}dr,\label{indfunc}
\end{equation}
with $|n|\geq 1$ and $\alpha>0$. $H$ is the completion of 
\begin{align}
X=\left\{u\in\mathcal{C}^1[0,R]| u(0)=0=u(R)\right\}
\end{align}
(the space of differentiable functions over $[0,R]$ which vanish at the two endpoints of the interval) and is equipped with the inner product 
\begin{align}
(u,\tilde{u})&=\int_0^R\left\{ru_r\tilde{u}_r+\dfrac{1}{r}u\tilde{u} \right\}dr,\quad u,\tilde{u}\in H.\label{innerPrd}
\end{align}
We may treat $H$ as an embedded subspace of $W_0^{1,2}(B_R)$, composed of radially symmetric functions enjoying the property $u(0)=0$ for any $u\in H$, where $B_R:=\left\{(x,y)\in\mathbb{R}^2:x^2+y^2\leq R^2\right\}$.

We are interested in finite energy solutions of \eqref{vortexEq}. From the inequality
\begin{align}
ln(1+x^2)\leq x^2\quad\text{for all $x\in\mathbb{R}$}\label{basic1}
\end{align} 
and 
\begin{align}
\int_{0}^{R}ru^2dr\leq R^2\int_{0}^{R}\dfrac{u^2}{r}dr\label{ineq1},
\end{align}
we note that the norm induced by the inner product \eqref{innerPrd} on $H$ guarantees that all terms in the functional $\mathcal{I}_{\kappa}$ stay finite. In other words, there is a constant $C>0$, such that $\mathcal{I}_{\kappa}(u)\leq C||u||_H^2.$
For convenience, we define the `energy' functional as
\begin{equation}
\mathcal{E}(u)=\dfrac{1}{2}\int_{0}^{R}\left\{ru^2_r+\frac{1}{r}u^2+r\ln(1+\alpha u^2)\right\}dr.\label{energy}
\end{equation}
\begin{theorem}
If $u$ is a nontrivial finite energy $(\mathcal{E}(u)<\infty)$ solution of the $n$-vortex equation \eqref{vortexEq}, then the wave propagation constant must satisfy
\begin{align}
\kappa<\alpha^{-1}-\dfrac{r_0^2+n^2}{2R^2},\label{necessarycond}
\end{align}
where $r_0(\approx 2.404825)$ is the first positive zero of the Bessel function $J_0$ \cite{YZ}.
\end{theorem}
\textbf{Proof.} Suppose if
$\liminf_{r\rightarrow 0}\left\{ru(r)|u_r(r)|\right\}\neq 0$, then there is an $\epsilon>0$ and $r_0\in(0,R]$ such that 
$ru(r)|u_r(r)|\geq \epsilon$ for all $r\in(0,r_0)$.
However, 
\begin{align}
\infty=\int_0^{r_0}\dfrac{\epsilon}{r}dr\leq\int_0^{r_0}u|u_r|dr\leq\left(\int_0^{r_0}\dfrac{u^2}{r}dr\right)^{1/2}\left(\int_0^{r_0}ru_r^2dr\right)^{1/2},
\end{align}
which contradicts the finite energy condition. Hence, 
$\liminf\limits_{r\rightarrow 0}\left\{ru(r)|u_r(r)|\right\}= 0$
and we can extract a sequence  $\{r_j\}$ such that $r_j\rightarrow 0$ as $j\rightarrow \infty$ and 
\begin{equation}
\liminf_{r\rightarrow 0}\left\{r_ju(r_j)|u_r(r_j)|\right\}=0.\label{2.12}
\end{equation}
Multiplying \eqref{vortexEq} by $u$, we integrate by parts over the interval $[r_j,R]$ and let $j\rightarrow\infty$. Appealing to \eqref{2.12}, we have
\begin{align}
-\int_{0}^{R}ru^2_rdr&=\int_{0}^{R}\left\{\frac{n^2}{r}u^2+2\kappa ru^2-\frac{2ru^4}{1+\alpha u^2}\right\}dr.\label{2.13}
\end{align}
Using $F(u)=\frac{u^2}{1+\alpha u^2}<\alpha^{-1}$ and \eqref{ineq1}, we get
\begin{align}
-\int_{0}^{R}ru^2_rdr&>-2\left(\alpha^{-1}-\dfrac{n^2}{2R^2}-\kappa\right)\int_{0}^{R}ru^2dr.
\end{align}
Treating $u$ as a radially symmetric function in $W_0^{1,2}(B_R)$ and using the Poincar\'e inequality 
\begin{align}
\int_0^R ru^2dr\leq \dfrac{R^2}{r_0^2}\int_0^R ru_r^2dr,
\end{align}
with $r_0$ as defined in \eqref{necessarycond}, gives
\begin{align}
0>-2\left(\alpha^{-1}-\dfrac{n^2+r_0^2}{2R^2}-\kappa\right)\int_{0}^{R}ru^2dr.\label{27}
\end{align}
Let 
\begin{equation}
\sigma:=\alpha^{-1}-\dfrac{n^2+r_0^2}{2R^2}-\kappa.\label{sigma}
\end{equation}
If $\sigma\leq 0$, then \eqref{27} gives a contradiction. Therefore, we must have $\sigma> 0$.$\qquad\square$

As a consequence of the following lemma, when the distance from the vortex core $R$ is sufficiently large, we prove that the functional \eqref{indfunc} is indefinite.
\begin{lemma} Let $\kappa<\alpha^{-1}$ and $\beta>0$. If the distance from the vortex core satisfies
\begin{align}
R>\left(\dfrac{12(1+n^2(2\ln 2 -1))}{\alpha^{-1}-\kappa}\right)^{\frac{1}{2}},\label{3.4}
\end{align}
then there exists an element $u_0\in H$ such that 
\begin{align}
||u_0||_H^2>\beta\qquad\text{and}\qquad \mathcal{I}_{\kappa}(u_0)< 0.
\end{align}
\end{lemma}
\textbf{Proof.} 
Set $R=2a$ and define
\begin{align}
u_0(r)=\left\{
\begin{array}{cc}
\frac{b}{a}r, &0\leq r\leq a,\\
\frac{b}{a}(2a-r), &a\leq r\leq 2a.
\end{array}\right.\label{2.28}
\end{align}
By direct calculation we obtain,
\begin{subequations}
\begin{align}
\int_0^{2a} ru_0^2dr&=\dfrac{2}{3}a^2b^2,\label{2.29}\\
\int_0^{2a} ru_{0,r}^2dr&=2b^2,\label{2.30}\\
\int_0^{2a} \dfrac{1}{r}u_0^2dr&=2b^2(2\ln 2-1),\label{2.31}\\
\int_0^{2a}r\ln(1+\alpha u_0^2)dr&=2a^2\left(\ln(1+\alpha b^2)-2+\dfrac{2}{\sqrt{\alpha}b}\tan^{-1}(\sqrt{\alpha }b)\right),\label{2.32}
\end{align}
\end{subequations}
where $u_{0,r}:=(u_0)_r$. Similar to Lemma 3.3 in \cite{YZ}, we note that $u_0$ is obtained as the limit of a Cauchy sequence in $H$, consequently, $u_0$ belongs in $H$. Using $(2.19)$, we get
\begin{align}
||u_0||_H^2&=4b^2\ln(2),\\
\mathcal{I}_{\kappa}(u_0)=& b^2\left(1+n^2(2\ln 2-1)-\dfrac{2}{3}(\alpha^{-1}-\kappa)a^2\right.\\
&+\left.\dfrac{2a^2}{\alpha^2b^2}\left[\ln(1+\alpha b^2)-2+\dfrac{2}{\sqrt{\alpha}b}\tan^{-1}(\sqrt{\alpha }b)\right]\right)\nonumber
\end{align}
with $a=R/2$. For any $\epsilon>0$, choose $b$ sufficiently large such that
\begin{align}
\dfrac{2}{\alpha^2b^2}\left|\ln(1+\alpha b^2)-2+\dfrac{2}{\sqrt{\alpha}b}\tan^{-1}(\sqrt{\alpha }b)\right|\leq\epsilon.\label{epsl}
\end{align}
Hence, 
\begin{align}
\mathcal{I}_{\kappa}(u_0)\leq b^2\left(1+n^2(2\ln 2-1)+\left(\epsilon-\dfrac{2}{3}(\alpha^{-1}-\kappa)\right)a^2\right).\label{3.8}
\end{align}
In order for the right hand term of \eqref{3.8} to be negative, let $\epsilon=\dfrac{1}{3}(\alpha^{-1}-\kappa)$ and $R$ be such that
\begin{align}
R>\left(\dfrac{12(1+n^2(2\ln 2 -1))}{\alpha^{-1}-\kappa}\right)^{\frac{1}{2}}.
\end{align}
For any $\beta>0$, choose $b$ satisfying \eqref{epsl} and, such that, $||u_0||_H^2>\beta$. With these values of $b$ and $R$, we have $\mathcal{I}_{\kappa}(u_0)<0$.$\qquad\square$

From inequality \eqref{3.8}, if the distance from the vortex core is sufficiently large, then $\mathcal{I}_{\kappa}(u_0)\rightarrow-\infty$ as $b\rightarrow\infty$ for $\kappa<\alpha^{-1}$. Therefore, for a sufficiently large distance $R$ and $\kappa<\alpha^{-1}$, the functional $\mathcal{I}_{\kappa}$ is indefinite and as such, a direct minimization is not possible.

% % % % % % % % % % % % % % % % % % % % % % % % % % % % % % % % % % % % % % % % % % % %
\section{Existence via constrained minimization}

Using a variational principle and constrained minimization problem, we prove the existence of positive solutions of the $n$-vortex equation \eqref{vortexEq}. In this scenario, the wave propagation constant $\kappa$ is undetermined and appears as a Lagrange multiplier. We provide a series of explicit estimates for the wave propagation constant, vortex winding number, and a prescribed energy flux. 

We view \eqref{vortexEq} as a nonlinear eigenvalue problem 
\begin{align}
\left\{\begin{array}{c}
(ru_r)_r-\dfrac{n^2}{r}u+\dfrac{2ru^3}{1+\alpha u^2}=2\kappa ru,\\
u(0)=0, \qquad u(R)=0.
\end{array}\right.\label{EigProb}
\end{align}
Define the action functional $\mathcal{I}$ as,
\begin{equation}
\mathcal{I}(u)=\frac{1}{2}\int_{0}^{R}\left\{ru^2_r+\frac{n^2}{r}u^2-2\alpha^{-1}ru^2+2\alpha^{-2}r\ln(1+\alpha u^2)\right\}dr,\label{2.3}
\end{equation}
and the soliton energy flux constraint functional $Q$ as
\begin{equation}
Q(u)=2\pi\int_{0}^{R}ru^2dr.\label{constraint}
\end{equation}
Consider the nonempty admissible class
\begin{equation}
\mathcal{A}=\left\{ \text{$u(r)$ is absolutely continuous over $[0,R]$}, u(0)=u(R)=0,\mathcal{E}(u)<\infty\right\},\label{admisClass}
\end{equation}
where $\mathcal{E}(u)$ is a defined by \eqref{energy}. In order to prove the existence of a solution pair $(u,\kappa)$, it suffices to show that a solution to the following exist:
\begin{equation}
\eta=\inf_{u\in\mathcal{A}}\left\{\mathcal{I}(u): Q(u)=Q_0>0\right\}\label{minProb},
\end{equation}
where $Q_0$ is a prescribed value for the energy flux and $\kappa$ is the Lagrange multiplier. 

\begin{theorem}
Consider the $n$-vortex equation \eqref{vortexEq}, describing ring-profile vortex solitons in a self-focusing saturable nonlinear media, subject to the prescribed energy flux $Q(u)=Q_0>0$ and finite-energy condition $\mathcal{E}(u)<\infty$, defined by \eqref{constraint} and \eqref{energy}, respectively, with parameters $|n|\geq 1$, $\alpha>0$, and $R>0$.
\begin{enumerate}[(i)]
\item There exists a solution pair $(u,\kappa)$  satisfying $u(r)>0$ for $r\in(0,R)$.
\item If $n^2+2r^2\kappa >0$ for $r\in[0,R]$, then there exists no nontrivial small-energy-flux solution satisfying $Q(u)=Q_0\leq 1/4$.
\end{enumerate}
\end{theorem}
\textbf{Proof.}
$(i)$ From the prescribed energy flux, it follows that the functional $\mathcal{I}(u)$ satisfies
\begin{equation}
\mathcal{I}(u)\geq\frac{1}{2}\int\limits_{0}^{R}ru^2_rdr+n^2\int\limits_{0}^{R}\dfrac{u^2}{r}dr-\dfrac{\alpha^{-1}}{2\pi}Q_0.\label{coercBound}
\end{equation} 
As a result, the minimization problem \eqref{minProb} is well-defined. Let $\{u_j\}$ be a minimizing sequence of \eqref{minProb}, i.e., choose a sequence of functions $\{u_j\}$ in $\mathcal{A}$ such that 
\begin{align*}
\mathcal{I}(u_j)\rightarrow \eta\qquad\text{as}\qquad j\rightarrow\infty\qquad\text{and}\qquad \mathcal{I}(u_1)\geq\mathcal{I}(u_2)\geq\ldots\geq\eta.
\end{align*}
Since $\{u_j\}$ minimizes \eqref{minProb} and using \eqref{coercBound}, there exist $C>0$ independent of $j$ such that 
\begin{equation}
C\geq\int_{0}^{R}ru_{j,r}^2dr+\int_{0}^{R}\dfrac{u_j^2}{r}dr,\label{2.16}
\end{equation}
where $u_{j,r}:=\frac{d}{dr}u_j$. 

The distributional derivative of $u$ satisfies $||u|_r|\leq|u_r|$, and the functionals $\mathcal{I}$ and $Q$ are both even, i.e., $\mathcal{I}(u)=\mathcal{I}(|u|)$ and $Q(u)=Q(|u|)$. As a consequence, we assume that the sequence  $\{u_j\}$ consists of nonnegative valued functions. Moreover, we take these functions to be radially symmetric over the disk $B_R$ and vanishing on its boundary.

From \eqref{2.16} and  \eqref{ineq1} it can be seen that the functions $u_j$ belong in $W^{1,2}_0(B_R)$ under the radially symmetric reduced norm,
\begin{align}
||u||^2:=\int_{0}^{R}ru^2 dr+\int_{0}^{R}ru_r^2 dr.\nonumber
\end{align}
Using \eqref{2.16} and \eqref{ineq1}, we show the sequence $\{u_j\}$ is bounded in $W_0^{1,2}(B_R)$, 
\begin{align}
||u_j||^2=\int_{0}^{R}ru_j^2dr+\int_{0}^{R}ru_{j,r}^2dr\leq R^2\int_{0}^{R}\dfrac{u_j^2}{r}dr+\int_{0}^{R}ru_{j,r}^2dr\leq (R^2+1)C.\nonumber
\end{align}
Without loss of generality, since we are in a reflexive space, we may assume the weak convergence of $\{u_j\}$ to an element $u\in W^{1,2}_0(B_R)$. As a result, it now suffices to show that $u_j$ converges to a minimizer of \eqref{minProb} and belongs in $\mathcal{A}$. 

From the compact embedding $W^{1,2}(B_R)\subset\subset L^p(B_R)$ for $p\geq 1$, $u_j\rightarrow u$ strongly in $L^p(B_R)$ as $j\rightarrow\infty$. Hence, $u$ is radially symmetric and satisfies the boundary condition $u(R)=0$. 

In view of \eqref{2.16} and using Fatou's lemma, we get
\begin{subequations}
\begin{align}
\int_0^R ru^2_rdr\leq \liminf_{j\rightarrow\infty}\int_0^Rru^2_{j,r}dr,\label{Fatou1}\\
\int_0^R \frac{u^2}{r}dr\leq \liminf_{j\rightarrow\infty}\int_0^R \frac{u^2_j}{r}dr,\\
\int_0^R r\ln(1+\alpha u^2)dr\leq \liminf_{j\rightarrow\infty}\int_0^Rr\ln(1+\alpha u_j^2)dr,\label{Fatou3}
\end{align}
\end{subequations}
where the finiteness of the right hand side of \eqref{Fatou3} follows from \eqref{basic1} and \eqref{ineq1}. Therefore, from \eqref{ineq1} and $(3.8)$, we get the weak lower semi-continuity of the functional $\mathcal{I}$, i.e.,
\begin{align}
\mathcal{I}(u)\leq\liminf_{j\rightarrow\infty}\mathcal{I}(u_j).\label{lowersemi}
\end{align}
Using \eqref{lowersemi}, together with \eqref{minProb}, gives $\mathcal{I}(u)=\eta$. Further, note that $Q(u)=\lim\limits_{j\rightarrow\infty}Q(u_j)=Q_0$. Moreover, the finite-energy condition also holds from \eqref{2.16} and \eqref{Fatou1}-\eqref{Fatou3}. In particular, $ru^2$, $\frac{u^2}{r}$, $ru_r^2$, and $\ln(1+\alpha u^2)$ are all in $L(0,R)$. 

To show that $u(0)=0$, we follow as in \cite{YZ}. Let $\{u_j\}$ be a sequence in $W^{1,2}(\epsilon,R)$ where $\epsilon\in(0,R)$. For any $\epsilon\in(0,R)$, $\{u_j\}$ is bounded in $W^{1,2}(\epsilon,R)$. The compact embedding $W^{1,2}(\epsilon,R)\subset\subset C[\epsilon,R]$, then gives $u_j\rightarrow u$ uniformly over $[\epsilon,R]$ as $j\rightarrow\infty$. Thus, for any pair $r_1,r_2\in (0,R)$ such that $r_1<r_2$ and using $C$ from \eqref{2.16}, we get
\begin{align}
|u_j^2(r_2)-u_j^2(r_1)|&=\left|\int_{r_1}^{r_2}(u^2_j(r))_rdr\right|\\
&\leq\int_{r_1}^{r_2}2|u_j(r)u_{j,r}(r)|dr\nonumber\\
&\leq 2 \left(\int_{r_1}^{r_2}ru_{j,r}^2(r)dr\right)^{1/2}\left(\int_{r_1}^{r_2}\dfrac{u_j^2(r)}{r}dr\right)^{1/2}\nonumber\\
&\leq 2 C^{1/2}\left(\int_{r_1}^{r_2}\dfrac{u_j^2(r)}{r}dr\right)^{1/2}.\nonumber
\end{align}
Since $u_j\rightarrow u$ uniformly as $j\rightarrow\infty$, we take $j\rightarrow \infty$ above to get
\begin{align}
|u^2(r_2)-u^2(r_1)|\leq 2 C^{1/2}\left(\int_{r_1}^{r_2}\dfrac{u^2(r)}{r}dr\right)^{1/2}.\label{2.22}
\end{align}
The right hand side of \eqref{2.22} goes to zero as $r_1,r_2\rightarrow 0$; since $\frac{u^2}{r}$ is in $L(0,R)$. Hence, the following limit exists,
\begin{equation}
\xi_0=\lim_{r\rightarrow 0}u^2(r)=0.
\end{equation}
As a consequence, the boundary condition $u(0)=0$ is achieved. 

Therefore, the function $u$, obtained as the limit of the minimizing sequence $\{u_j\}$, is a solution to the constrained minimization problem \eqref{minProb}, and there is a real number $\kappa$ such that $(u,\kappa)$ satisfies \eqref{EigProb}. 

Further, we may suppose that there is a point $r_0\in(0,R)$ such that $u(r_0)=0$. Since $r_0$ would be a minimum point for $u(r)$, we have $u_r(r_0)=0$. However, by the uniqueness theorem of the initial value problem of ordinary differential equations, $u(r)=0$ for all $r\in(0,R)$, thus contradicting the energy flux constraint $Q(u)=Q_0>0$. Hence, $u(r)>0$ for all $r\in(0,R)$. A standard bootstrap method may then be used to conclude that $u$ is a classical solution of \eqref{vortexEq}.

$(ii)$ Let $(u,\kappa)$ be the solution pair obtained in part $(i)$. Using $\frac{u^4}{1+\alpha u^2}\leq u^4$ in \eqref{2.13}, we get
\begin{align}
-\int_{0}^{R}ru^2_rdr&\geq\int_0^R\left(\dfrac{n^2}{r^2}+2\kappa\right)ru^2dr-2\int_0^R ru^4dr.
\end{align}
We treat $u$ as a radially symmetric function defined over $\mathbb{R}^2$ with its support contained in the disk $B_R$. From the classical Gagliardo-Nirenberg inequality over $\mathbb{R}^2$, we write
\begin{align}
\int_0^R ru^4dr\leq 4\pi\int_0^R ru^2dr\int_0^R ru^2_rdr.
\end{align}
As a result, we get
\begin{align}
-\int_{0}^{R}ru^2_rdr&\geq\int_0^R\left(\dfrac{n^2}{r^2}+2\kappa\right)ru^2dr-8\pi\int_0^R ru^2dr\int_0^R ru^2_rdr.\label{above}
\end{align}
Rearranging the terms in \eqref{above}, and using the prescribed energy flux constraint gives
\begin{align}
\left(4Q_0-1\right)\int_0^R ru^2_rdr-\int_0^R\left(\dfrac{n^2}{r^2}+2\kappa\right)ru^2dr\geq 0.\label{56}
\end{align}
Hence, $u\equiv 0$, if 
\begin{align}
Q_0\leq\dfrac{1}{4}\quad\text{and}\quad \dfrac{n^2}{r^2}+2\kappa >0\quad \text{for}\quad r\in(0,R],
\end{align} 
as claimed.$\qquad\square$

\begin{theorem}
Let $(u,\kappa)$ be the solution pair of the $n$-vortex equation \eqref{vortexEq} obtained in Theorem 3.1 with $\kappa$ as the wave propagation constant.
\begin{enumerate}[(i)]
\item The wave propagation constant satisfies
\begin{align}
\kappa\geq& \alpha^{-1} -\dfrac{6 }{R^2}\left(1+n^2(2\ln 2-1)\right)\\
&-\dfrac{\pi R^2}{\alpha^2 Q_0}\left[\ln\left(1+\dfrac{3\alpha Q_0}{\pi R^2}\right)-2+\sqrt{\dfrac{4\pi R^2}{3\alpha Q_0}}\tan^{-1}\left(\sqrt{\dfrac{3\alpha Q_0}{\pi R^2}}\right)\right]\nonumber.
\end{align}
\item If the vortex winding number satisfies $|n|\geq Q_0/\pi$, then $\kappa<0$.
\item For $\kappa>0$, the solution pair $(u,\kappa)$ satisfies
\begin{equation*}
u^2\leq C_{\kappa}\exp(-\sqrt{2\kappa}r),
\end{equation*}
for $r$ sufficiently large and $C_{\kappa}>0$ is a constant depending on $\kappa$ only.
\end{enumerate}
\end{theorem}
\textbf{Proof.}
$(i)$ To obtain a lower bound for $\kappa$, we rearrange \eqref{2.13} and write
\begin{align}
\kappa\int_{0}^{R}ru^2dr=-\dfrac{1}{2}\int_{0}^{R}\left(ru^2_r+\frac{n^2}{r}u^2\right)dr+\int_0^R\frac{ru^4}{1+\alpha u^2}dr.\label{2.24}
\end{align}
Choose $u_0\in\mathcal{A}$, satisfying $Q(u_0)=Q_0>0$. Since $u$ is a solution to the constrained minimization problem \eqref{minProb}, whose existence was proved in Theorem 3.1, we have $\mathcal{I}(u)\leq \mathcal{I}(u_0)$. As such, we get
\begin{align}
\dfrac{1}{2}\int_0^R\left(ru^2_r+\dfrac{n^2}{r}u^2\right)dr\leq \mathcal{I}(u_0)+\alpha^{-1}\int_0^R ru^2 dr-\alpha^{-2}\int_0^Rr\ln(1+\alpha u^2)dr.\nonumber
\end{align}
Inserting the above into \eqref{2.24} and using $Q(u)=Q_0>0$, gives
\begin{align}
\kappa&\geq-\alpha^{-1}-\dfrac{2\pi}{Q_0}\mathcal{I}(u_0)+\alpha^{-2}\dfrac{2\pi}{Q_0}\int_0^Rr\ln(1+\alpha u^2)dr+\dfrac{2\pi}{Q_0}\int_0^R\frac{ru^4}{1+\alpha u^2}dr.\label{3.20}
\end{align}
Using the inequality 
\begin{align}
\ln(1+x)\geq \frac{x}{1+x}\quad\text{ for $x\geq 0$,} \label{basic2}
\end{align}
\eqref{basic2} may be rewritten as
\begin{align}
\kappa&\geq-\alpha^{-1}-\dfrac{2\pi}{Q_0}\mathcal{I}(u_0)+\alpha^{-1}\dfrac{2\pi}{Q_0}\int_0^R r\frac{u^2+\alpha u^4}{1+\alpha u^2}dr=-\dfrac{2\pi}{Q_0}\mathcal{I}(u_0).\label{2.27}
\end{align}
From \eqref{2.28} and $(2.19)$, we obtain
\begin{align}
\mathcal{I}(u_0)&=b^2+n^2b^2(2\ln 2-1)-\dfrac{2}{3}\alpha^{-1}a^2b^2\label{2.33}\\
&+\dfrac{2a^2}{\alpha^2}\left[\ln(1+\alpha b^2)-2+\dfrac{2\tan^{-1}(\sqrt{\alpha }b)}{\sqrt{\alpha}b}\right].\nonumber
\end{align}
Using \eqref{2.29} and $Q(u_0)=Q_0$, we get $b^2=3Q_0/(\pi R^2)$. Inserting \eqref{2.33} in \eqref{2.27}, we arrive at
\begin{align}
\kappa\geq& \alpha^{-1} -\dfrac{6 }{R^2}\left(1+n^2(2\ln 2-1)\right)\\\nonumber
&-\dfrac{3}{\alpha^2 b^2}\left[\ln\left(1+\alpha b^2\right)-2+\dfrac{2}{\sqrt{\alpha}b}\tan^{-1}\left(\sqrt{\alpha}b\right)\right],\nonumber
\end{align}
which is the desired lower bound. 

$(ii)$ Let $(u,\kappa)$ be the solution pair obtained in Theorem 3.1. Using Schwartz's inequality, and $u(0)=0$, we get
\begin{align}
u^2(r)=\int_0^r 2u(\rho)u_{\rho}(\rho)d\rho\leq 2\left(\int_0^R\rho u_\rho^2(\rho)d\rho\right)^{1/2}\left(\int_0^R\dfrac{u^2(\rho)}{\rho}d\rho\right)^{1/2}.\label{40}
\end{align}
Multiplying \eqref{40} by $ru^2$, integrating from $0$ to $R$, and using the constraint $Q(u)=Q_0>0$, we obtain
\begin{align}
\int_0^R ru^4dr\leq \dfrac{Q_0}{\pi}\left(\int_0^R\rho u_\rho^2(\rho)d\rho\right)^{1/2}\left(\int_0^R\dfrac{u^2(\rho)}{\rho}d\rho\right)^{1/2}.
\end{align}
Using inequality, $ab\leq \epsilon a^2+\frac{b^2}{4\epsilon}$ for every $a,b\in\mathbb{R}$ and $\epsilon>0$, gives
\begin{align}
\int_0^R ru^4dr\leq \epsilon\int_0^R\rho u_\rho^2(\rho)d\rho+\dfrac{Q_0^2}{\epsilon 4\pi^2}\int_0^R\dfrac{u^2(\rho)}{\rho}d\rho.\label{42}
\end{align}
From \eqref{42} and $\frac{u^4}{1+\alpha u^2}\leq u^4$ in \eqref{2.24}, we get
\begin{align}
\kappa \int_{0}^{R}ru^2dr&\leq-\left(\dfrac{1}{2}-\epsilon\right)\int_0^R ru_r^2 dr-\left(\dfrac{n^2}{2}-\dfrac{Q_0^2}{4\epsilon\pi^2}\right)\int_{0}^{R}\dfrac{u^2}{r}dr.\label{43}
\end{align}
We choose $\epsilon=\frac{1}{2}$ in \eqref{43} and conclude that $\kappa< 0$ whenever $|n|\geq Q_0/\pi$.

$(iii)$ The exponential decay estimate follows from an application of the maximum principle and a suitable exponential comparison function. We rewrite \eqref{EigProb} as
\begin{align}
\Delta u=u_{rr}+\dfrac{1}{r}u_r=\left(\dfrac{n^2}{r^2}-\dfrac{2u^2}{1+\alpha u^2}+2\kappa\right)u.\label{2.35}
\end{align}
It then follows,
\begin{align}
\Delta u^2\geq 2u\Delta u=2\left(\dfrac{n^2}{r^2}-\dfrac{2u^2}{1+\alpha u^2}+2\kappa\right)u^2\geq 4\left(\kappa-\dfrac{u^2}{1+\alpha u^2}\right)u^2. 
\end{align}
By the continuity of $u$ on $[0,R]$ and the boundary condition $u(R)=0$, for any $\epsilon>0$ there is an $R_{\epsilon}>0$ such that 
\begin{align}
\Delta u^2\geq 4\left(\kappa-\epsilon\right)u^2\quad\text{for every $r\in[R_{\epsilon},R]$.}\label{2.37}
\end{align}
Define the comparison function $\xi:[0,R]\rightarrow\mathbb{R}$ as
\begin{equation}
\xi(r)=Ce^{-\sigma r},\quad C,\sigma>0.\label{2.38}
\end{equation}
Hence,
\begin{equation}
\Delta\xi=\sigma^2\xi-\dfrac{\sigma\xi}{r}.\label{2.39}
\end{equation}
Subtracting \eqref{2.37} and \eqref{2.39}, for every $r\in [R_{\epsilon},R]$, 
\begin{align}
\Delta (u^2-\xi)&\geq 4\left(\kappa-\epsilon\right)u^2-\left(\sigma^2\xi-\dfrac{\sigma\xi}{r}\right)\geq 4\left(\kappa-\epsilon\right)u^2-\sigma^2\xi.
\end{align}
For any $\kappa>0$, we choose $\epsilon$ satisfying $0<\epsilon<\kappa$ and $\sigma^2=4(\kappa-\epsilon)$, to get 
\begin{equation}
\Delta (u^2-\xi)\geq \sigma^2(u^2-\xi)\quad\text{for every}\quad r\in [R_{\epsilon},R].
\end{equation}
Let $C$ in \eqref{2.38} large enough so that $u^2-\xi\leq 0$ for $r=R_{\epsilon}$. Denote $C$ by $C_{\epsilon}$ to emphasize its dependence on $\epsilon$. Since, $u^2\rightarrow 0$ as $r\rightarrow R^-$, and applying the maximum principle, we conclude that  $u^2-\xi\leq 0$ for all $r\in[R_{\epsilon},R]$. For simplicity, let $\epsilon=\kappa/2$ to obtain
\begin{align}
u^2\leq C_{\kappa}\exp(-\sqrt{2\kappa}r)\quad\text{for every $r\in[R_{\kappa},R]$},
\end{align}
where $C_{\kappa}>0$, and $R_{\kappa}$ depends only on $\kappa>0$.$\qquad\square$

\textbf{Remarks.} Beam confinement requires the exponential decay of the soliton amplitude $u$ at infinity \cite{SF}. From the exponential decay estimate given in Theorem 3.2 $(iii)$, we see that this occurs for $\kappa>0$. Theorem 3.2 $(ii)$ states that the vortex winding number must satisfy $|n|<Q_0/\pi$. On the other hand, from Theorem 3.1 $(ii)$, if the wave propagation constant is positive, then the prescribed energy flux must satisfy $Q_0>\frac{1}{4}$. As will be seen in Section 6, the condition on the prescribed energy flux, $Q_0>\frac{1}{4}$, is not sufficient to conclude that the propagation constant is positive. 

When the prescribed energy flux $Q_0$ is fixed, and the distance from the vortex core $R$ goes to infinity, the necessary condition given in Theorem 2.1, together with the lower bound for the wave propagation constant, gives the inequality
\begin{align}
0<\kappa<\alpha^{-1},\label{2.7}
\end{align}
which is in agreement with the results of Skryabin and Firth \cite{SF}, for any self-trapped solutions of their model.

% % % % % % % % % % % % % % % % % % % % % % % % % % % % % % % % % %
% % % % % % % % % % % % % % % % % % % % % % % % % % % % % % % % % %
\section{Solutions on the Nehari Manifold}
Recall the action functional $\mathcal{I}_{\kappa}:H\rightarrow\mathbb{R}$ defined as,
\begin{equation}
\mathcal{I}_{\kappa}(u)=\frac{1}{2}\int_{0}^{R}\left\{ru^2_r+\frac{n^2}{r}u^2-2(\alpha^{-1}-\kappa)ru^2+2\alpha^{-2}r\ln(1+\alpha u^2)\right\}dr,\label{3.1}
\end{equation}
where $|n|\geq 1$ and $\alpha>0$.

Standard arguments show that $\mathcal{I}_{\kappa}\in\mathcal{C}^3(H,\mathbb{R})$. Also,
\begin{align}
\langle\mathcal{I}'_{\kappa}(u),\tilde{u}\rangle=\int_{0}^{R}\left\{ru_r\tilde{u}_r+\frac{n^2}{r}u\tilde{u}-2(\alpha^{-1}-\kappa)ru\tilde{u}+2\alpha^{-1}\dfrac{ru\tilde{u}}{1+\alpha u^2}\right\}dr,\qquad \forall \tilde{u}\in H,\label{4.2}
\end{align}
and $\langle\cdot{,}\cdot\rangle$ denotes the usual duality between $H$ and its dual space $H^{-1}$. Let $\gamma_{\kappa}:H\rightarrow\mathbb{R}$ be defined by 
\begin{align}
\gamma_{\kappa}(u)=\dfrac{1}{2}\langle\mathcal{I}'_{\kappa}(u),u\rangle=\dfrac{1}{2}\int_{0}^{R}\left\{ru_r^2+\frac{n^2}{r}u^2-2(\alpha^{-1}-\kappa)ru^2+2\alpha^{-1}\dfrac{ru^2}{1+\alpha u^2}\right\}dr.
\end{align}
For a fixed propagation constant $\kappa$, define the Nehari manifold $\mathcal{M}$ as
\begin{align}
\mathcal{M}=\{u\in H\backslash \{0\}:\gamma_{\kappa}(u)=0\}.\label{Nehari}
\end{align} If $u\in H\backslash \{0\}$ is a critical point of $\mathcal{I}_{\kappa}$, then for every $v\in H$, $\langle\mathcal{I}_{\kappa}(u),v\rangle=0$. Setting $v=u$, it follows that $\gamma_{\kappa}(u)=\frac{1}{2}\langle\mathcal{I}_{\kappa}(u),u\rangle=0$.  Hence, $u\in\mathcal{M}$, and the Nehari manifold contains all nontrivial critical points of $\mathcal{I}_{\kappa}$ on $H$.

\begin{lemma}
$u\in H$ is a nontrivial critical point of $\mathcal{I}_{\kappa}$ if and only if $u\in\mathcal{M}$ and is a critical point of $\mathcal{I}_{\kappa}|_{\mathcal{M}}$. 
\end{lemma}
\textbf{Proof.} The forward implication follows directly from the definition of the Nehari manifold. We now justify the other direction. For every $u\in\mathcal{M}$,
\begin{align}
\int_{0}^{R}\left\{ru_r^2+\frac{n^2}{r}u^2-2(\alpha^{-1}-\kappa)ru^2\right\}dr=-2\alpha^{-1}\int_0^R\left\{\dfrac{ru^2}{1+\alpha u^2}\right\}dr.
\end{align}
By definition of $\gamma_{\kappa}$, for every $u\in\mathcal{M}$, we get
\begin{align}
\langle\gamma_{\kappa}'(u),u\rangle&=\int_{0}^{R}\left\{ru_r^2+\frac{n^2}{r}u^2-2(\alpha^{-1}-\kappa)ru^2+2\alpha^{-1}\dfrac{ru^2}{(1+\alpha u^2)^2}\right\}dr\nonumber\\
&=2\alpha^{-1}\int_{0}^{R}\left\{\dfrac{ru^2}{1+\alpha u^2}\left(\dfrac{1}{1+\alpha u^2}-1\right)\right\}dr<0.\label{3.7}
\end{align}
For any critical point $u_o\in\mathcal{M}$ of $\mathcal{I}_{\kappa}|_{\mathcal{M}}$, there exists a Lagrange multiplier, $\xi\in\mathbb{R}$, such that $\langle\mathcal{I}'_{\kappa}(u_0),\tilde{u}\rangle=\xi\langle\gamma_{\kappa}'(u_0),\tilde{u}\rangle$, for every $\tilde{u}\in H$. Using the definition of $\mathcal{M}$, gives
\begin{align}
0=\gamma_{\kappa}(u_0)=\dfrac{1}{2}\langle\mathcal{I}'_{\kappa}(u_0),u_0\rangle=\xi\langle\gamma_{\kappa}'(u_0),u_0\rangle. 
\end{align}
As a result, using \eqref{3.7}, it follows that $\xi=0$. Therefore, the critical points of $\mathcal{I}_{\kappa}|_{\mathcal{M}}$ are also the critical points of $\mathcal{I}_{\kappa}$. $\quad\square$

Lemma 4.1 indicates that $\mathcal{M}$ is a natural constraint for $\mathcal{I}_{\kappa}$. From the necessary condition obtained in Theorem 2.1, the Nehari manifold contains no critical points of $\mathcal{I}_{\kappa}$ when
$\kappa\geq\alpha^{-1}-\frac{r_0^2+n^2}{2R^2}$. Hence, we consider the case $\kappa<\alpha^{-1}-\frac{r_0^2+n^2}{2R^2}$.
\begin{lemma}
If the distance from the vortex core satisfies
\begin{align}
\left(\dfrac{6(1+n^2(2\ln(2)-1))}{\alpha^{-1}-\kappa}\right)^{\frac{1}{2}}<R,\label{kRnempty}
\end{align}
then the Nehari manifold is not empty.
\end{lemma}
\textbf{Proof.} Define 
\begin{align}
\Gamma(t,u)=\dfrac{1}{2}\int_{0}^{R}\left\{ru_r^2+\frac{n^2}{r}u^2-2(\alpha^{-1}-\kappa)ru^2+2\alpha^{-1}\dfrac{ru^2}{1+\alpha t^2u^2}\right\}dr.
\end{align}
As a result, $\gamma_{\kappa}(tu)=t^2\Gamma(t,u)$. For any $u\neq 0$ and $\kappa>-\frac{r_0^2+n^2}{2R^2}$, we get
\begin{align}
\Gamma(0,u)=\dfrac{1}{2}\int_{0}^{R}\left\{ru_r^2+\frac{n^2}{r}u^2+2\kappa ru^2\right\}dr\geq\left(\dfrac{r_0^2+n^2}{2R^2}+\kappa\right)\int_0^Rru^2dr>0.
\end{align}
Note that
\begin{align}
\Gamma(\infty,u)=\lim\limits_{t\rightarrow\infty}\Gamma(t,u)=\dfrac{1}{2}\int_{0}^{R}\left\{ru_r^2+\frac{n^2}{r}u^2-2(\alpha^{-1}-\kappa)ru^2\right\}dr.\label{Gam}
\end{align}
Substituting $u_0$ as defined in Lemma 2.2, in \eqref{Gam}, we get
\begin{align}
\Gamma(\infty,u_0)=b^2\left(1+n^2(2\ln(2)-1)-\dfrac{1}{6}(\alpha^{-1}-\kappa)R^2\right). 
\end{align}
Selecting $R$ as in \eqref{kRnempty}, we get $\Gamma(\infty,u_0)<0$. Hence, there exists a $t_0>0$ such that $\Gamma(t_0,u_0)=0$ and, it follows  that, $\gamma_{\kappa}(t_0u_0)=t_0^2\Gamma(t_0,u_0)=0$. Therefore, $t_0u_0$ is in the Nehari manifold $\mathcal{M}$.
$\quad\square$

Note that, for every $\kappa>-\frac{r_0^2+n^2}{2R^2}$,
\begin{align}
\langle\gamma_{\kappa}''(0),\tilde{u}\rangle=\int_0^R\left\{r\tilde{u}_r^2+\dfrac{n^2}{r}\tilde{u}^2+2\kappa r\tilde{u}^2\right\}dr\geq 2\left(\dfrac{r_0^2+n^2}{2R^2}+\kappa\right)\int_0^R r\tilde{u}^2dr>0.
\end{align}
Hence, $u=0$ is a strict local minimum of $\gamma_{\kappa}$ and, as a result, an isolated point in $\mathcal{M}\cup\{0\}$. Thus, $0\notin\partial\mathcal{M}$. Therefore, for all $u\in\mathcal{M}$, there exists a constant $C_1>0$, independent of $u$, such that 
\begin{align}
||u||_H\geq C_1\label{c1}.
\end{align}
\begin{lemma}
There exists a constant $C_2>0$, such that 
\begin{align}
\langle\gamma_{\kappa}'(u),u\rangle<-C_2<0\qquad\text{for all $u\in\mathcal{M}$}.\label{c2}
\end{align}
\end{lemma}
\textbf{Proof.}
Let $\{u_j\}_{j=1}^{\infty}$ be a sequence in $\mathcal{M}$ such that 
\begin{align}
\lim\limits_{j\rightarrow\infty}\int_{0}^{R}\dfrac{ru_j^2}{1+\alpha u_j^2}dr=0. \label{3.12}
\end{align} 
Hence,  $\{u_j\}_{j=1}^{\infty}$ is either bounded or unbounded in $H$. If  $\{u_j\}_{j=1}^{\infty}$ is unbounded, then there is a subsequence  $\{u_{j_k}\}_{k=1}^{\infty}$, such that $\rho_k:=||u_{j_k}||_H\rightarrow\infty$ as $k\rightarrow\infty$. Let $v_k=\frac{u_{j_k}}{\rho_k}$, and hence $||v_k||_H=1$. If necessary, passing to a subsequence, $v_k\rightharpoonup v$ in $H$, $v_k\rightarrow v$ in $L^2(B_R)$, and $v_k(r)\rightarrow v(r)$ a.e. $r\in[0,R]$ as $k\rightarrow\infty$. Consequently, 
\begin{align}
0=\lim\limits_{k\rightarrow\infty}\int_{0}^{R}\dfrac{ru_{j_k}^2}{1+\alpha u_{j_k}^2}dr=\lim\limits_{k\rightarrow\infty}\int_{0}^{R}\dfrac{rv_k^2}{\frac{1}{\rho_k}+\alpha v_k^2}dr=\dfrac{1}{2}\alpha^{-1}R^2>0,
\end{align}
which is a contradiction. Hence, $\{u_j\}_{j=1}^{\infty}$ must be a bounded sequence in $H$. If necessary, passing to a subsequence, we have $u_j\rightharpoonup u$ in $H$, $u_j\rightarrow u$ in $L^2(B_R)$, and $u_j(r)\rightarrow u(r)$ a.e. $r\in[0,R]$ as $j\rightarrow\infty$.
Using the dominated convergence theorem, we get
\begin{align}
0=\lim\limits_{j\rightarrow\infty}\int_0^R\dfrac{ru_j^2}{1+\alpha u_j^2}dr=\int_0^R\dfrac{ru^2}{1+\alpha u^2}dr,
\end{align}
which gives $u(r)=0$ a.e. $r\in[0,R]$. Since $\{u_j\}_{j=1}^{\infty}$ is in $\mathcal{M}$, we get
\begin{align}
\lim\limits_{j\rightarrow\infty}\int_{0}^{R}\left\{ru_{j,r}^2+\frac{n^2}{r}u_j^2\right\}dr&=\lim\limits_{j\rightarrow\infty}\left\{(\alpha^{-1}-\kappa)\int_{0}^{R}ru_j^2dr-2\alpha^{-1}\int_{0}^{R}\dfrac{ru_j^2}{1+\alpha u_j^2}dr\right\}\nonumber\\&=0.
\end{align}
It follows for any $|n|\geq 1$,
\begin{align}
0=\lim\limits_{j\rightarrow\infty}\int_{0}^{R}\left\{ru_{j,r}^2+\frac{n^2}{r}u_j^2\right\}dr\geq\lim\limits_{j\rightarrow\infty}||u_j||_H^2\geq 0.
\end{align}
Hence, $u_j\rightarrow 0$ in $H$, which contradicts \eqref{c1}. Therefore, there exists a constant $C_3>0$ such that 
\begin{align}
\int_{0}^{R}\dfrac{ru^2}{1+\alpha u^2}dr\geq C_3\quad\text{for every $u\in\mathcal{M}$.}
\end{align} 
Using \eqref{3.7}, and Holder's inequality, we get
\begin{align}
\langle\gamma_{\kappa}'(u),u\rangle&=-2\int_{0}^{R}\dfrac{ru^4}{(1+\alpha u^2)^2}dr\\
&\leq-\dfrac{4}{R^2}\left(\int_{0}^{R}\dfrac{ru^2}{1+\alpha u^2}dr\right)^2\nonumber\\
&\leq-\dfrac{4}{R^2}C_3^2=: -C_2<0.\nonumber
\end{align}
$\quad\square$
\begin{lemma}
The set $\mathcal{M}$ is a paracompact and complete topological space. 
\end{lemma}
\textbf{Proof.} The paracompactness of the set $\mathcal{M}$ follows identically to the proof in Lemma 3.6, \cite{LiuRen}. To show that $\mathcal{M}$ is complete, we let $\{u_j\}_{j=1}^{\infty}$ be a sequence in $\mathcal{M}$ such that $u_j\rightarrow u$ in $H$ as $j\rightarrow\infty$. From the compact embedding, $W^{1,2}(B_R)\subset\subset L^p(B_R)$ for $p\geq 1$, we note that $u_j\rightarrow u$ strongly in $L^p(B_R)$ as $j\rightarrow\infty.$ Hence,
\begin{align}
0=\lim\limits_{j\rightarrow\infty}\gamma_{\kappa}(u_j)&=\dfrac{1}{2}\int_{0}^{R}\left\{ru_r^2+\frac{n^2}{r}u^2-2(\alpha^{-1}-\kappa)ru^2+2\alpha^{-1}\dfrac{ru^2}{1+\alpha u^2}\right\}dr\\
&=\gamma_{\kappa}(u).\nonumber
\end{align}
Since $||u||_H\geq C_1>0$, by \eqref{c1}, we conclude that $u\in\mathcal{M}$. $\quad\square$

Similarly to \cite{LiuRen}, using Lemma 4.3, we may deduce that $\mathcal{M}$ is a regular $\mathcal{C}^2$-Banach manifold and, moreover, using Lemma 4.4 that $\mathcal{M}$ is a Finsler manifold. We now look for nontrivial solutions of the $n$-vortex equation \eqref{vortexEq}, as critical points of $\mathcal{I}_{\kappa}$ restricted to the manifold $\mathcal{M}$. 
\begin{lemma}
Let $\{u_j\}_{j=1}^{\infty}$ be a sequence in $\mathcal{M}$ such that $\{\mathcal{I}_{\kappa}(u_j)\}$ is bounded. Then the sequence $\{u_j\}_{j=1}^{\infty}$ is bounded in $H$.
\end{lemma}
\textbf{Proof.}
Let $\{u_j\}_{j=1}^\infty$ be a sequence in $\mathcal{M}$ such that $\{\mathcal{I}_{\kappa}(u_j)\}$ is bounded. Hence,
\begin{align}
0=\gamma_{\kappa}(u_j)=\dfrac{1}{2}\int_0^R\left\{ru_{j,r}^2+\dfrac{n^2}{r}u_j^2-2(\alpha^{-1}-\kappa)ru_j^2+2\alpha^{-1}\dfrac{ru_j^2}{1+\alpha u_j^2}\right\},\label{psd}
\end{align}
and there exists a constant $\beta>0$ independent of $j$ such that
\begin{align}
|\mathcal{I}_{\kappa}(u_j)|=\left|\frac{1}{2}\int_{0}^{R}\left\{ru^2_{j,r}+\frac{n^2}{r}u_j^2-2(\alpha^{-1}-\kappa)ru_j^2+2\alpha^{-2}r\ln(1+\alpha u_j^2)\right\}dr\right|\leq\beta.\label{psb}
\end{align}
Using \eqref{psd} and \eqref{psb}, we get
\begin{align}
\int_{0}^{R}\left\{\ln(1+\alpha u_j^2)-\dfrac{\alpha u_j^2}{1+\alpha u_j^2}\right\}rdr\leq\alpha^2\beta.\label{bnd}
\end{align}
Assume the sequence $\{u_j\}_{j=1}^{\infty}$ is unbounded in $H$. Let $\rho_j=||u_j||_H$. Then, $\rho_j\rightarrow\infty$ as $j\rightarrow\infty$. Let $v_j=\frac{u_j}{\rho_j}$. Then $||v_j||_H=1$. Hence, without loss of generality, we suppose that $v_j\rightharpoonup v$ in $H$, $v_j\rightarrow v$ in $L^p(B_R)$ for every $p\geq 1$, and $v_j(r)\rightarrow v(r)$ a.e. $r\in[0,R]$. From \eqref{psd},
\begin{align}
1=||v_j||_H^2&\leq\int_0^R\left\{rv_{j,r}^2+\dfrac{n^2}{r}v_j^2\right\}dr
=\int_0^R\left\{2(\alpha^{-1}-\kappa)rv_j^2-2\alpha^{-1}\dfrac{rv_j^2}{1+\alpha u_j^2}\right\}dr\nonumber\\
&\leq\int_0^R\left\{2(\alpha^{-1}-\kappa)rv_j^2\right\}dr=2(\alpha^{-1}-\kappa)||v_j||_{L^2(B_R)}^2.\label{4.21}
\end{align}
Letting $j\rightarrow\infty$ in \eqref{4.21}, we get $1\leq 2(\alpha^{-1}-\kappa)||v||_{L^2(B_R)}^2$. Hence, $v\not\equiv 0$ a.e. $r\in[0,R]$. Let $\Omega=\{r\in[0,R]:v(r)\neq 0\}$. Then $|\Omega|\neq 0$. Using \eqref{basic2}, it follows
\begin{align}
\int_{0}^{R}\left\{\ln(1+\alpha u_j^2)-\dfrac{\alpha u_j^2}{1+\alpha u_j^2}\right\}rdr\geq\int_{\Omega}\left\{\ln(1+\alpha u_j^2)-\dfrac{\alpha u_j^2}{1+\alpha u_j^2}\right\}rdr.
\end{align}
Note that as a result $v_j(r)=\frac{u_j(r)}{\rho_j}\rightarrow v(r)\neq 0$ a.e. $r\in\Omega$ as $j\rightarrow\infty$. Hence, $|u_j(r)|\rightarrow\infty$ and $\ln(1+\alpha u_j(r)^2)-\dfrac{\alpha u_j(r)^2}{1+\alpha u_j(r)^2}\rightarrow\infty$ a.e. $r\in\Omega$. Applying Fatou's lemma and \eqref{bnd}, we get the contradiction
\begin{align}
\alpha^2\beta&\geq\liminf\limits_{j\rightarrow\infty}\int_{0}^{R}\left\{\ln(1+\alpha u_j^2)-\dfrac{\alpha u_j^2}{1+\alpha u_j^2}\right\}rdr\nonumber\\
&\geq\int_{\Omega}\liminf\limits_{j\rightarrow\infty}\left\{\ln(1+\alpha u_j^2)-\dfrac{\alpha u_j^2}{1+\alpha u_j^2}\right\}rdr=\infty.
\end{align}
Therefore, the sequence $\{u_j\}_{j=1}^{\infty}$ is bounded in $H$.$\quad\square$

\begin{lemma}
$\mathcal{I}_{\kappa}$ satisfies the Palais-Smale condition on $\mathcal{M}$, namely, if $\{u_j\}_{j=1}^{\infty}$ is a sequence in $\mathcal{M}$ such that $\{\mathcal{I}_{\kappa}(u_j)\}$ is bounded and $\mathcal{I}_{\kappa}|'_{\mathcal{M}}(u_j)\rightarrow 0$, then there exists a $u\in\mathcal{M}$ such that $u_j\rightarrow u$  (strongly) in $H$. Moreover, $u$ is a critical point of $\mathcal{I}_{\kappa}|_{\mathcal{M}}$.
\end{lemma}
\textbf{Proof.} Suppose that $\{\mathcal{I}_{\kappa}(u_j)\}$ is bounded. Then, from Lemma 4.5, $\{u_j\}_{j=1}^{\infty}$ is bounded in $H$. Without loss of generality, there is a sequence $\{u_j\}_{j=1}^{\infty}$ such that $u_j\rightharpoonup u$ in $H$, $u_j\rightarrow u$ in $L^p(B_R)$ for every $p\geq 1$, and $u_j(r)\rightarrow u(r)$ a.e. $r\in[0,R]$. For every $v\in H$, we have
\begin{align}
\langle\mathcal{I}'_{\kappa}(u_j),v\rangle&=\int_{0}^{R}\left\{ru_{j,r}v_r+\frac{n^2}{r}u_jv-2(\alpha^{-1}-\kappa)ru_jv+2\alpha^{-1}\dfrac{ru_jv}{1+\alpha u_j^2}\right\}dr\nonumber\\
&\rightarrow\int_{0}^{R}\left\{ru_{r}v_r+\frac{n^2}{r}uv-2(\alpha^{-1}-\kappa)ruv+2\alpha^{-1}\dfrac{ruv}{1+\alpha u^2}\right\}dr\nonumber\\
&=\langle\mathcal{I}'_{\kappa}(u),v\rangle
\end{align}
and
\begin{align}
\langle\gamma_{\kappa}'(u_j),v\rangle&=\int_{0}^{R}\left\{ru_{j,r}v+\frac{n^2}{r}u_jv-2(\alpha^{-1}-\kappa)ru_jv+2\alpha^{-1}\dfrac{ru_jv}{(1+\alpha u_j^2)^2}\right\}dr\nonumber\\
&\rightarrow\int_{0}^{R}\left\{ru_rv_r+\frac{n^2}{r}uv-2(\alpha^{-1}-\kappa)ruv+2\alpha^{-1}\dfrac{ruv}{(1+\alpha u^2)^2}\right\}dr\nonumber\\
&=\langle\gamma_{\kappa}'(u),v\rangle.
\end{align}
Using the definition of $\mathcal{I}_{\kappa}|'_{\mathcal{M}}$, there exists a sequence $\{\xi_j\}_{j=1}^{\infty}$ in $\mathbb{R}$ such that 
\begin{align}
\mathcal{I}_{\kappa}'(u_j)-\xi_j\gamma_{\kappa}'(u_j)\rightarrow 0\quad\text{in $H^{-1}$ as $j\rightarrow\infty$}.\label{4.26}
\end{align}
If $\{\xi_j\}_{j=1}^{\infty}$ is unbounded, then there exists a renamed subsequence $\{\xi_j\}_{j=1}^{\infty}$ such that $\xi_j\rightarrow\infty$. From \eqref{4.26}, $\xi_j\langle\gamma_{\kappa}'(u_j),u\rangle\rightarrow\langle\mathcal{I}_{\kappa}'(u),u\rangle$ and, hence, $\langle\gamma_{\kappa}'(u_j),u\rangle\rightarrow 0$. As a result, $\langle\gamma_{\kappa}'(u),u\rangle=0$. However, using \eqref{c2}, we get 
\begin{align}
0>-C_2\geq\lim\limits_{j\rightarrow\infty}\langle\gamma_{\kappa}'(u_j),u_j\rangle\geq\langle\gamma_{\kappa}'(u),u\rangle=0,
\end{align}
a contradiction. Therefore, $\{\xi_j\}_{j=1}^{\infty}$ is bounded and contains a renamed subsequence $\{\xi_j\}_{j=1}^{\infty}$, such that $\xi_j\rightarrow\xi$ as $j\rightarrow\infty$. Consequently, \eqref{4.26} implies
\begin{align}
\mathcal{I}_{\kappa}'(u)-\xi\gamma_{\kappa}'(u)=0.\label{4.33}
\end{align}
Suppose $\xi=1$. From \eqref{4.33},
\begin{align}
0=\langle\mathcal{I}_{\kappa}'(u)-\gamma_{\kappa}'(u), u\rangle=2\alpha^{-1}\int_0^R\left\{\dfrac{ru^2}{1+\alpha u^2}\left(1-\dfrac{1}{1+\alpha u^2}\right)\right\}dr,
\end{align}
which implies that $u(r)=0$ a.e. $r\in[0,R]$. Thus,
\begin{align}
0=\gamma_{\kappa}(u_j)&=\dfrac{1}{2}\langle\mathcal{I}'_{\kappa}(u_j),u_j\rangle\\
&=\dfrac{1}{2}\int_0^R\left\{ru_{j,r}^2+\dfrac{n^2}{r}u_j^2-2(\alpha^{-1}-\kappa)ru_j^2+2\alpha^{-1}\dfrac{ru_j^2}{1+\alpha u_j^2}\right\}dr,\nonumber
\end{align}
and using the dominated convergence theorem, we conclude that $||u_j||^2_H\rightarrow 0$, which is a contradiction to \eqref{c1}. Therefore, $\xi\neq 1$.

The boundedness of $\{u_j\}_{j=1}^{\infty}$ and \eqref{4.26}, gives
\begin{align}
|\langle\mathcal{I}_{\kappa}'(u_j)-\xi_j\gamma_{\kappa}'(u_j),u_j-u\rangle|\leq||\mathcal{I}_{\kappa}'(u_j)-\xi_j\gamma_{\kappa}'(u_j)||_{H^{-1}}||u_j-u||_H\rightarrow 0.\label{4.37}
\end{align}
then applying \eqref{4.33}, we have
\begin{align}
\langle\mathcal{I}_{\kappa}'(u_j)-\xi_j\gamma_{\kappa}'(u_j),u_j-u\rangle=&\langle\mathcal{I}_{\kappa}'(u_j)-\xi_j\gamma_{\kappa}'(u_j)-\mathcal{I}_{\kappa}'(u)+\xi\gamma_{\kappa}'(u),u_j-u\rangle\nonumber\\
=&\langle\mathcal{I}_{\kappa}'(u_j)-\mathcal{I}_{\kappa}'(u),u_j-u\rangle-\langle\xi_j\gamma_{\kappa}'(u_j)-\xi\gamma_{\kappa}'(u),u_j-u\rangle\nonumber\\
=&\langle\mathcal{I}_{\kappa}'(u_j)-\mathcal{I}_{\kappa}'(u),u_j-u\rangle-\xi_j\langle\gamma_{\kappa}'(u_j)-\gamma_{\kappa}'(u),u_j-u\rangle\nonumber\\
&-(\xi_j-\xi)\langle\gamma_{\kappa}'(u),u_j-u\rangle.
\end{align}
Using the definition of $\mathcal{I}_{\kappa}$ and $\gamma_{\kappa}$, we obtain
\begin{align}
\langle\mathcal{I}_{\kappa}'(u_j)-\mathcal{I}_{\kappa}'(u),u_j-u\rangle=&\int_{0}^{R}\left\{r(u_{j,r}-u_r)^2+\frac{n^2}{r}(u_j-u)^2-2(\alpha^{-1}-\kappa)r(u_j-u)^2\right\}dr\nonumber\\
&+2\alpha^{-1}\int_0^R\left\{\left(\dfrac{u_j}{1+\alpha u_j^2}-\dfrac{u}{1+\alpha u_j^2}\right)(u_j-u)\right\}rdr
\end{align}
and
\begin{align}
\langle\gamma_{\kappa}'(u_j)-\gamma_{\kappa}'(u),u_j-u\rangle=&\int_{0}^{R}\left\{r(u_{j,r}-u_r)^2+\frac{n^2}{r}(u_j-u)^2-2(\alpha^{-1}-\kappa)r(u_j-u)^2\right\}dr\nonumber\\
&+2\alpha^{-1}\int_0^R\left\{\left(\dfrac{u_j}{(1+\alpha u_j^2)^2}-\dfrac{u}{(1+\alpha u_j^2)^2}\right)(u_j-u)\right\}rdr.
\end{align}
Applying the dominated convergence theorem, we get
\begin{align}
\int_0^R\left\{\left(\dfrac{u_j}{1+\alpha u_j^2}-\dfrac{u}{1+\alpha u_j^2}\right)(u_j-u)\right\}rdr\rightarrow 0
\end{align}
and 
\begin{align}
\int_0^R\left\{\left(\dfrac{u_j}{(1+\alpha u_j^2)^2}-\dfrac{u}{(1+\alpha u_j^2)^2}\right)(u_j-u)\right\}rdr\rightarrow 0.\label{4.40}
\end{align}
Since $u_j\rightharpoonup u$ in $H$, $u_j\rightarrow u$ in $L^p(B_R)$ for every $p\geq 1$, and $u_j(r)\rightarrow u(r)$ a.e. $r\in[0,R]$, from equations \eqref{4.37}-\eqref{4.40} and \eqref{ineq1}, we conclude
\begin{align}
(1-\xi)\int_0^Rr(u_{j,r}-u_r)^2dr\rightarrow 0.
\end{align}
Therefore, $||u_j-u||_H\rightarrow 0$ in $H$. From the completeness of $\mathcal{M}$, Lemma 4.4, $u\in\mathcal{M}$. Then, \eqref{4.33} implies that $u$ is a critical point of $\mathcal{I}_{\kappa}|_{\mathcal{M}}$. $\quad\square$
\begin{theorem}
Let the distance from the vortex core satisfy \eqref{kRnempty}. For each propagation constant in the interval
\begin{align}
\left(-\dfrac{n^2+r_0^2}{2R^2}, \alpha^{-1}-\dfrac{n^2+r_0^2}{2R^2}\right),
\end{align}
there exists a solution pair $(u,\kappa)$, satisfying $u(r)>0$ for $r\in(0,R)$, to the $n$-vortex equation \eqref{vortexEq}.
\end{theorem}
\textbf{Proof.}
Let $u\in\mathcal{M}$. Hence $\gamma_{\kappa}(u)=0$ and 
\begin{align}
\int_{0}^{R}\left\{ru_r^2+\frac{n^2}{r}u^2\right\}dr=\int_0^R\left\{2(\alpha^{-1}-\kappa)ru^2-2\alpha^{-1}\dfrac{ru^2}{1+\alpha u^2}\right\}dr.\label{4.34}
\end{align}
Inserting \eqref{4.34} into $\mathcal{I}_{\kappa}$ and using \eqref{basic2}, gives
\begin{align}
\mathcal{I}_{\kappa}(u)=\alpha^{-2}\int_0^R\left\{\ln(1+\alpha u^2)-\dfrac{\alpha u^2}{1+\alpha u^2}\right\}rdr>0\quad \text{on $\mathcal{M}$}.\label{defOnM}
\end{align}
Thus the functional $\mathcal{I}_{\kappa}$ is bounded below on $\mathcal{M}$. As a consequence of Ekeland's variatonal principle \cite{Ekel, Jabri} and Lemma 4.6, there exists a $u\in\mathcal{M}$ such that $\mathcal{I}_{\kappa}(u)=\inf\{\mathcal{I}_{\kappa}(v)|v\in\mathcal{M}\}$ and $\mathcal{I}_{\kappa}|_{\mathcal{M}}'(u)=0$. By Lemma 4.1, $u$ is also a critical point of $\mathcal{I}_{\kappa}$ and, therefore, a solution of the $n$-vortex equation \eqref{vortexEq}. 

We use the evenness of the functional $\mathcal{I}_{\kappa}$ to get a positive solution. Moreover, $u(r)>0$ for all $r\in(0,R)$. Suppose there is a point $r_0\in(0,R)$ such that $u(r_0)=0$. Then $r_0$ would be a minimum point for $u(r)$ and $u_r(r_0)=0$. By the uniqueness theorem of the initial value problem of ordinary differential equations, $u(r)=0$ for all $r\in(0,R)$, thus contradicting the fact that $u\in\mathcal{M}$. $\quad\square$

\section{Finite Element Formalism}
We utilize the variational principle used in Section 3 and a finite element formalism to compute the solution pair $(u,\kappa)$ to the problem \eqref{vortexEq}, for a prescribed energy flux \eqref{energyFlux}. This is essentially achieved by approximating the solutions to the constrained minimization problem \eqref{minProb}. 

Recall the admissible class $\mathcal{A}$, defined in \eqref{admisClass}. Let $V$ be a subset of $\mathcal{A}$, composed of $N$ linearly independent functions, $\left\{\psi_j\right\}_{j=1}^N$. Define the inner product as
\begin{align}
(u,\tilde{u})=2\pi\int_0^Rru\tilde{u}dr,\qquad u,\tilde{u}\in\mathcal{A},\label{5.1}
\end{align}
whose form is suggested by the constraint functional \eqref{constraint}. Under the inner product \eqref{5.1}, the set $V$ can be orthonormalized via the Gram-Schmidt procedure. We let the functions $\left\{\psi_j\right\}_{j=1}^N$ in $V$ be orthonormal with respect to the inner product \eqref{5.1}. We approximate functions $u\in \mathcal{A}$ by using the finite element formalism
\begin{align}
u=\sum_{j=1}^{N}a_j\psi_j,\label{5.2}
\end{align} 
with $a_1,\ldots,a_N\in\mathbb{R}$. Using this formalism \eqref{5.2}, the constrained minimization problem \eqref{minProb} becomes
\begin{align}
\min\left\{F(a)=\mathcal{I}\left(\sum_{j=1}^{N}a_j\psi_j\right)\quad\bigg|\quad\sum_{j=1}^{N}a_j^2=Q_0,\quad a\in\mathbb{R}^N\right\},\label{5.3}
\end{align}
where $a=(a_1,\ldots,a_N),$ is called the variational vector. Note that $F$ is a continuous, real-valued function defined over the surface of the $N$-sphere of radius $\sqrt{Q_0}$ centered at the origin in $\mathbb{R}^N$. Hence, the constrained minimization problem \eqref{5.3} is well-defined and has a nontrivial solution. 

We use MATLAB's Optimization Toolbox \cite{Matlab} and the Chebfun package \cite{chebfun} to solve  \eqref{5.3}. In particular, we obtain a minimum using the objective function as
\begin{align}
F(a)=\dfrac{1}{2}\sum^{N}_{i,j=1}a_ia_j\int_0^R\Biggl\{r\psi_{i,r}\psi_{j,r}+\dfrac{n^2}{r}\psi_i\psi_j-\dfrac{2\alpha^{-1}r^2\psi_i\psi_{j,r}}{1+\alpha\biggl(\sum\limits_{k=1}^{N}a_k\psi_k\biggl)^2}\Biggl\}dr-\dfrac{Q_0}{2\alpha\pi}.\nonumber
\end{align}
In order to compute the wave propagation constant, we use a Lagrange multiplier $\lambda\in\mathbb{R}$, such that $\langle\mathcal{I}(u),\tilde{u}\rangle=\lambda \langle(u),\tilde{u}\rangle$. More explicitly, there exists a $\lambda\in\mathbb{R}$ such that
\begin{equation}
\int_0^R\left\{ru_r\tilde{u}_r+\dfrac{n^2}{r}u \tilde{u}-\dfrac{2ru^3\tilde{u}}{1+\alpha u^2}\right\}dr=4\pi\lambda\int_0^Rru\tilde{u}dr\label{5.6}
\end{equation}
for every $\tilde{u}\in \mathcal{A}$. Comparing the weak formulation of the $n$-vortex equation \eqref{vortexEq}, we get $\kappa = -2\pi\lambda$. Using $\tilde{u}=u$ in \eqref{5.6} and the prescribed energy flux $Q_0>0$, gives
\begin{align}
\kappa =-\dfrac{\pi}{Q_0}\int_0^R\left\{ ru_r^2+\frac{n^2}{r}u^2 -\frac{2ru^4}{1+\alpha u^2}\right\}dr.\label{5.7}
\end{align}

As in \cite{SF}, we consider the case when the saturation constant, vortex winding number, and distance from the vortex core are:
$\alpha = 0.1$, $n=1$, and $R=8$, respectively. With this particular choice of parameter values and using Theorem 2.1 and Theorem 3.2$(iii)$, we get
\begin{equation}
0<\kappa<9.9470.\label{5.8}
\end{equation} 
Equivalently, using \eqref{5.7}, we get that the prescribed energy flux satisfies
\begin{equation}
13.6<Q_0<\infty. \label{5.9}
\end{equation}
The inequalities \eqref{5.8} and \eqref{5.9} are the necessary conditions for postive exponentially decaying solutions. This in turn, numerically demonstrates that $Q_0>\frac{1}{4}$, as remarked in Section 3, does imply that $\kappa>0$. 
\begin{figure*}[h]
\centering
\includegraphics[scale=0.35]{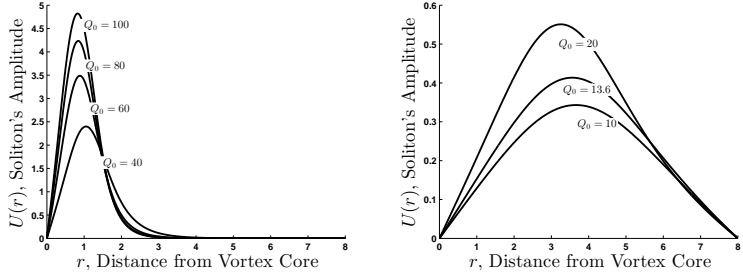}
\caption{Soliton's amplitude for $\alpha=$ $0.1$, $n=1$, $R=8$, and $N=40$.}
\label{fig:fig1}
\end{figure*}

Figure \ref{fig:fig1}, shows the soliton's amplitude for several values of the energy flux $Q_0$. Note that, as the prescribed energy flux $Q_0$ is increased the soliton's amplitude also increases. The numerical error is estimated by substituting the formalism \eqref{5.2} into
\begin{equation}
error=\int_0^R\left((ru_r)_r-\dfrac{n^2}{r}u+2r\dfrac{u^3}{1+\alpha u^2}-2\kappa ru\right)^2 dr.
\end{equation}

For $Q_0 = 40, 60, 80, 100$, we compute the propagation constant $\kappa$ and obtain $\kappa = 1.4901$, $2.5827$, $3.2955$, $3.8120$ with $error = 0.0001, 0.0050, 0.0120$, $0.0116$, respectively. Figure \ref{fig:fig1}, also illustrates the behavior of the soliton's amplitude for the borderline values of $Q_0=10, 13.6, 20$. The following values for the propagation constant $\kappa=-.0330, 0.0001,0.0712$ with $error=0.0170,0.0178,0.0156$, respectively, are obtained. As expected, the value of the wave propagation constant is negative when $Q_0<13.6$.

%We see that our computation is in agreement with that of Skryabin and Firth in \cite{SF} for $n=1$, but begins to deviate as $n$ increases. The qualitative behaviour of $\kappa$ is also in agreement for large $Q_0$ as well. It would be interesting to further explore the relationship between $Q_0$ and $\kappa$.
\begin{figure*}[h]
\centering
\includegraphics[scale=0.4]{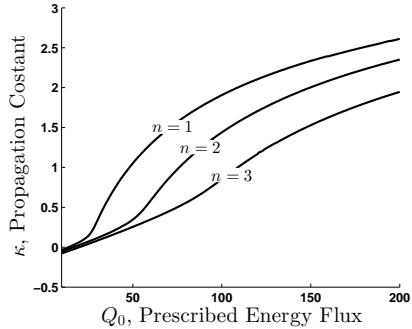}
\caption{Propagation constant $\kappa$ as a function of the prescribed energy flux $Q_0$ with fixed parameter values $\alpha=0.2$, $R=8$, and $N=15$ for $n=1,2,3$.}
\label{fig:fig2}
\end{figure*}

We also analyze the behavior of the solution pair $(u,\kappa)$ for a fixed value of the energy flux $Q_0$ by varying the vortex winding number $n$ (see Figure \ref{fig:fig2}). Particularly, when $\alpha=0.1$ and $R=8$, Theorem 2.1 states that the wave propagation constant $\kappa$ must satisfy $\kappa<10-(r_0^2+n^2)/128$, which imposes an upper limit on the vortex winding number of $|n|<\sqrt{1280-r_0^2}\approx 35.6962$ (i.e., $|n|\leq 35$). However, for exponentially decaying solutions, i.e., $\kappa>0$, owing to Theorem 3.2$(ii)$, the vortex winding number is bounded above by $Q_0/\pi$. Consequently and for example, using $Q_0 = 10\pi$, we get  $|n|<10$  as a necessary condition for positive exponentially decaying solutions. 

\begin{figure}[h]
\centering
\includegraphics[scale=0.3]{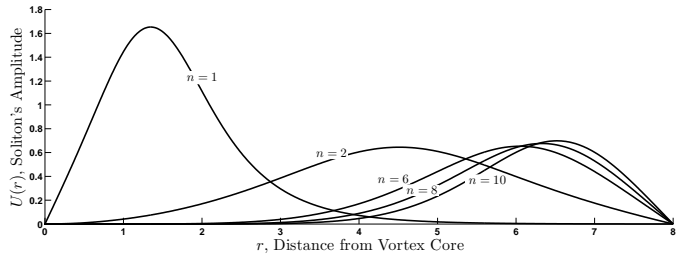}
\caption{Soliton's amplitude for $n = 1,2,6,8,10$ with $Q_0=10\pi$, $\alpha=0.1$, $R=8$, and $N=20$.} 
\label{fig:fig3}
\end{figure}

Figure \ref{fig:fig3} shows the values for the wave propagation constant $\kappa= 0.7933$, $0.0607$, $-0.4812$, $-0.8562$, $-1.3046$ for $n = 1,2,6,8,10$ with $Q_0=10\pi$. We observe that the wave propagation constant decreases as the vortex winding number increases, which is expected and implied from the necessary condition of Theorem 2.1. In particularly, $\kappa\rightarrow -\infty$ as $n\rightarrow\infty$.

We remark that our numerical approach is in contrast with that of Skryabin and Firth \cite{SF}. We compute the wave propagation constant for a prescribed energy flux (see Figure \ref{fig:fig2}). On the other hand, Skryabin and Firth in \cite{SF}, compute the soliton's amplitude for a prescribed propagation constant and then use \eqref{energyFlux} to determine its corresponding energy flux. 

\section{Conclusion}
In this paper, we establish a series of existence results for ring-profiled localized optical vortex solitons. We consider such solitons in the context of an electromagnetic wave propagating in a saturable nonlinear medium and model by a nonlinear Schr\"odinger equation \eqref{NLSeq}. In particular, we focus on spatially localized ring-profiled optical vortex solitons governed by the $n$-vortex equation \eqref{vortexEq}. Below we summarize the results:
\begin{enumerate}
\item From Theorem 2.1 and Theorem 3.2, a necessary condition for the existence of positive exponentially decaying solutions of the $n$-vortex equation \eqref{vortexEq} is
\begin{align}
0<\kappa<\alpha^{-1}-\dfrac{n^2+r_0^2}{2R^2}.\label{6.1}
\end{align}
Moreover, the vortex winding number must satisfy $|n|<Q_0/\pi$ (see Theorem 3.2$(ii)$) and the prescribed energy flux $Q_0>1/4$ (see Theorem 3.1$(ii)$). Further, no small-energy-flux solutions exists for $\kappa>0$ when $Q_0\leq 1/4$ (see Theorem 3.1$(ii)$).
\item The existence of a positive solution is guaranteed by Theorem 3.1$(i)$, however, the propagation constant $\kappa$ is undetermined. A lower bound for $\kappa$ is provided by Theorem 3.2$(i)$, and an upper bound by Theorem 2.1.
\item On a Nehari manifold, if the distance from the vortex core $R$ is sufficiently large, then for any propagation constant satisfying \eqref{6.1}, a positive exponentially decaying solution exists (see Theorem 4.7 and Theorem 3.2$(iii)$).

\item Using a finite element formalism, we compute the soliton's amplitude and wave propagation constant for a prescribed energy flux. The numerical analysis shows that the wave propagation constant increases as the energy flux increases and decreases as the vortex winding number increases. Moreover, for given parameter values $\alpha$, $n$, and $R$, we are able to numerically obtain a necessary condition for the existence of positive exponentially decaying solutions in terms of a prescribed energy flux (see \eqref{5.9}).
\end{enumerate}

\begin{acknowledgements}
	We thank Deane Yang, Yisong Yang, and Marie-Ange Brumelot for all the helpful conversations, as well as, the referee for the careful reading of the manuscript.  
\end{acknowledgements}

\def\bibsection{\section*{\refname}}

\end{document}